\documentclass[aps,prl,reprint,superscriptaddress,nofootinbib,longbibliography]{revtex4-2}

\usepackage{amsmath,amssymb}
\usepackage{graphicx}
\usepackage{bm}

\newcommand{\CC}{\mathbb{C}}
\newcommand{\Jac}{\operatorname{Jac}}

\newcommand{\SL}{\mathrm{SL}}
\newcommand{\SO}{\mathrm{SO}}
\newcommand{\gform}{G}
\newcommand{\id}{\mathbb{I}}

\begin{document}

\title{An entanglement band structure for general multiqubit states}

\author{Chon-Fai Kam}
\email{dubussygauss@gmail.com}
\affiliation{Dipartimento di Fisica e Chimica ``Emilio Segr\`e'',
Universit\`a degli Studi di Palermo, Via Archirafi 36, I-90123 Palermo, Italy}
\affiliation{Universit\'e Paris Cit\'e et Universit\'e de La R\'eunion,
75015 Paris, France}

\date{\today}

\begin{abstract}
Majorana drew a symmetric multiqubit state as points on the Bloch sphere,
antipodal to the zeros of its overlap with spin coherent states.
Projecting only one qubit of a general state leaves the others in a state
that varies with direction.
Its entanglement invariants are eigenvalues of a non-Hermitian matrix and
form bands over the sphere.
Exceptional points appear, without gain or loss, where the entanglement
family of the rest changes.
Circling one exchanges two bands, so the bands form one Riemann surface.
At five qubits it fixes the entanglement class up to finitely many choices.
\end{abstract}

\maketitle

Three qubits can be genuinely entangled in two inequivalent ways, GHZ and
W.
Two states of the same kind convert into each other by local operations and
classical communication with nonzero probability
(SLOCC)~\cite{DurVidalCirac2000}.
From four qubits on, the inequivalent kinds form continuous
families~\cite{Verstraete2002}, with a number of parameters that grows
exponentially in the number of qubits.
Polynomial invariants of growing degree tell these families
apart~\cite{Luque2006,Burchardt2024}.
They decide whether two states are equivalent, but give no picture of how
the states differ.
Entanglement polytopes, built from one-qubit spectra, are finitely many and
so coarse-grain the continuous families~\cite{Walter2013,Sawicki2014}.

For symmetric states a complete picture
exists~\cite{Zimba2006,Giraud2008,Romero2024,SanchezSoto2026}.
Majorana showed that a symmetric state of $n$ qubits is fixed, up to
normalization, by $n$ points on the Bloch
sphere~\cite{Majorana1932,Bacry1974}.
They are the directions antipodal to the zeros of
$\langle\bm n^{\otimes n}|\Psi\rangle$.
An invertible operation applied equally to every qubit moves the points by a
M\"obius map, so classifying symmetric states becomes classifying point
sets~\cite{Bastin2009,Burchardt2022}.
A symmetry of the state is a symmetry of the points, a degeneracy is a
coincidence of points, and the geometric phase follows from the solid angles
the points sweep~\cite{KamLiu2021}.
None of this reaches general states.
Their generic classes depend on $2^{n}-1-3n$ complex parameters, three at
four qubits and sixteen at five, while $n$ points on the sphere up to
M\"obius maps carry only $n-3$.
For three qubits, local operations bring every state into the symmetric
subspace while keeping its three-tangle, which then becomes a function of the
chordal distances between three stars~\cite{Kam2020}.
For four qubits this works only under special conditions.

\begin{figure*}[t]
  \includegraphics[width=\textwidth]{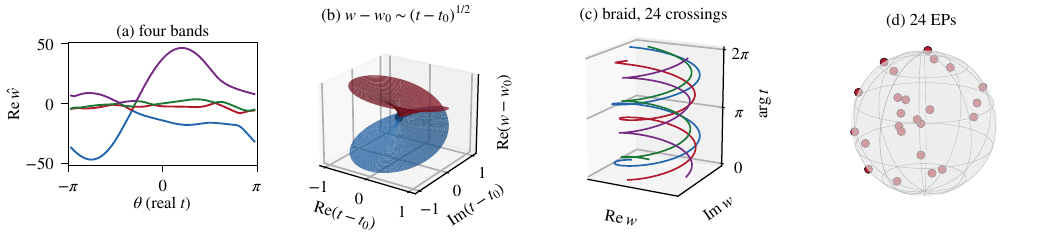}
  \caption{%
    Entanglement bands of the state $\Psi_{0}$ of the Supplemental
    Material~\cite{SM}, with the fifth qubit projected and the split $12|34$.
    (a) Real parts of the four bands along the great circle of real $t$,
    divided by $1+t^{2}$.
    No exceptional point lies on this circle, so every crossing is one of
    real parts only.
    (b) Near an exceptional point $t_{0}$, where two bands meet at $w_{0}$,
    they form a square-root sheet, with $t-t_{0}$ in units of the disc
    radius.
    (c) Around a loop enclosing all exceptional points the bands braid with
    24 crossings of one sign, the $(4,8)$ torus link, shown after a radial
    rescaling that leaves the braid unchanged.
    (d) The 24 exceptional points on the Bloch sphere.}
  \label{fig:one}
\end{figure*}

Here we extend Majorana's construction to general states by projecting
only one qubit.
Localizable entanglement optimizes over such measurements~\cite{Verstraete2004}.
We keep every measurement direction instead.
Majorana's points are the zeros of a holomorphic scalar, and projecting one
qubit instead gives a holomorphic matrix whose spectrum takes their place.
Projecting one qubit of a five-qubit state onto $|\bm n\rangle$ leaves a
four-qubit state.
For almost every $\bm n$ that state lies in the generic family $G_{abcd}$ of
Verstraete \textit{et al.}, labeled by four complex numbers
$a,b,c,d$~\cite{Verstraete2002}.
Their squares are invariants of the four-qubit state, and as $\bm n$ moves
over the Bloch sphere they vary like energy levels over a Brillouin zone and
form four entanglement bands (Fig.~\ref{fig:one}).
The reduced state of the four qubits is the same for every
$\bm n$~\cite{KamWong2026Wheeler}, so the bands follow the conditional
states alone.

Where the conditional state leaves the generic family, two bands meet at an
exceptional point (EP), and carrying $\bm n$ once around it exchanges them.
Exceptional points usually require gain, loss or coupling to an
environment~\cite{Ashida2020}.
Here they belong to a pure state and arise from conditioning alone, and
they can be encircled on a qubit processor by measuring Pauli operators.
The four bands are sheets of one compact Riemann surface, fixed by
the entanglement class of the five-qubit state.
The surface determines the class up to finitely many choices carried by the
eigenvectors.
In the same way the two phases of the Su--Schrieffer--Heeger chain share a
band spectrum and differ only in the Zak phase~\cite{SSH1979,Zak1989}.
Two five-qubit states are then compared through their band coefficients and
one linear system, with no search over local operations.
The construction works for any number of qubits, and that number sets a
non-Hermitian symmetry class.
Five is the only odd number above three at which the surface has as many
parameters as the classes.

\textit{Conditional states.}---Projecting a single qubit, instead of all of
them, leaves a state of the other qubits for every direction.
Let $t=\tan(\theta/2)e^{-i\varphi}$ be the stereographic coordinate of
$\bm n$ and $u=(u_{0},u_{1})=(1,t)$.
Projecting the fifth qubit onto $|\bm n\rangle$ gives the conditional state
$\cos(\theta/2)\,\psi_{u}$, whose amplitudes form the $4\times4$ matrix
$R(u)_{ab}=\sum_{s}\psi_{abs}u_{s}$.
Here $\psi_{abs}$ are the amplitudes of $|\Psi\rangle$, $a$ labels the basis
states of qubits 1 and 2, $b$ those of qubits 3 and 4, and $s$ those of the
fifth qubit.
$R$ is a pencil of matrices, linear in $u$, and holds each of the 32
amplitudes once.
The conditional state changes with $\bm n$, beyond normalization, exactly
when the fifth qubit is entangled with the rest.
For a two-qubit state $\phi$, the number $\phi^{\mathsf T}\gform\phi$ with
$\gform=\sigma_{y}\otimes\sigma_{y}$ is minus twice the determinant of its
coefficient matrix, and its modulus is the concurrence once $\phi$ is
normalized.
Local operations of unit determinant on a pair act as $\SO(4,\CC)$, which
preserves this form.
For almost every $\bm n$ the conditional state decomposes across the
$12|34$ split as
\begin{equation}
  |\psi_{u}\rangle=a\,|e_{1}\rangle|f_{1}\rangle+b\,|e_{2}\rangle|f_{2}\rangle
             +c\,|e_{3}\rangle|f_{3}\rangle+d\,|e_{4}\rangle|f_{4}\rangle ,
  \label{eq:bellschmidt}
\end{equation}
in frames with
$e_{i}^{\mathsf T}\gform e_{j}=f_{i}^{\mathsf T}\gform f_{j}=\delta_{ij}$
that are local images of the Bell basis.
Equation~\eqref{eq:bellschmidt} is the normal form of
$G_{abcd}$~\cite{Verstraete2002}, and
$a^{2},b^{2},c^{2},d^{2}$ are the eigenvalues of~\cite{Levay2006}
\begin{equation}
  L(u)=R(u)\,\gform\,R(u)^{\mathsf T}\gform .
  \label{eq:pencil}
\end{equation}
The coefficients enter squared because the frames are fixed only up to sign
and relabelling.
$L$ is quadratic in $u$, so each eigenvalue is homogeneous of degree two in
$u$, and these four functions are the bands.
The product $RR^{\dagger}$ would give the squared Schmidt coefficients
across the split, which local unitaries preserve but SLOCC does not.
The construction singles out one qubit to project and one way of splitting
the other four into two pairs.
Five choices of the qubit and three of the split give fifteen sets of bands
for a five-qubit state.
For a given projected qubit the three splits share the sum of the bands,
which up to sign is $\psi_{u}^{\mathsf T}\sigma_{y}^{\otimes4}\psi_{u}$.
With suitable signs, the determinants of $R$ for the three splits sum to
zero~\cite{LuqueThibon2003}.
The sphere plays the role of the Brillouin zone and $L(u)$ that of a
non-Hermitian Bloch Hamiltonian.

\begin{figure*}[t]
  \includegraphics[width=\textwidth]{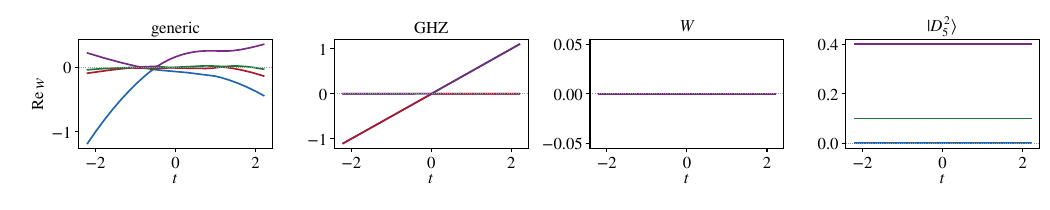}
  \caption{%
    Real parts of the bands along real $t$, without the division by
    $1+t^{2}$ of Fig.~\ref{fig:one}(a), for a generic five-qubit state and
    for the GHZ, $W$ and Dicke states.
    The generic bands braid into one surface, while the curves of the named
    states fall apart into copies of the sphere.}
  \label{fig:two}
\end{figure*}

\textit{Exceptional points.}---Two bands meet where two coefficients of
Eq.~\eqref{eq:bellschmidt} coincide up to sign.
There the frame of Eq.~\eqref{eq:bellschmidt} breaks down.
The frame vectors of the two meeting coefficients merge into a single
vector $e$ with $e^{\mathsf T}\gform e=0$.
It is a product state of qubits 1 and 2, and no rescaling brings it back to
$e^{\mathsf T}\gform e=1$.
The conditional state then lies in the family $L_{abc_{2}}$ of
Ref.~\cite{Verstraete2002} instead of $G_{abcd}$, and $L$ acquires a
$2\times2$ Jordan block.
The families of Ref.~\cite{Verstraete2002} are distinguished by the Jordan
form of $L$.
The exceptional points are therefore exactly the directions at which the
conditional state changes family.
A crossing at which $L$ stays diagonalizable leaves it in $G_{abcd}$.
These directions are the zeros of the discriminant of $L$, a polynomial of
degree 24 in $t$, so a generic five-qubit state has 24 of them, spread over
the sphere [Fig.~\ref{fig:one}(d)].
Hermitian bands would need three tuned parameters to meet, whereas these
holomorphic bands meet at isolated directions, as many as the degree of the
discriminant fixes.
At each of them the two bands share a single eigenvector.
Their difference grows as the square root of the distance from that
direction [Fig.~\ref{fig:one}(b)], the signature of an exceptional point of order
two~\cite{Berry2004,Heiss2012,Ashida2020}.
The number of qubits sets the symmetry class of $L$, and with it the kind of
degeneracy the bands can have.
Hermiticity is replaced by $L^{\mathsf T}\gform=\gform L$, which with
$\Gamma=\gform^{-\mathsf T}$ reads $\Gamma L^{\mathsf T}\Gamma^{-1}=L$ at every
direction, the transposed time-reversal symmetry of the 38-fold
classification of non-Hermitian matrices~\cite{Kawabata2019}.
For $n\equiv1\pmod4$, $\Gamma\Gamma^{*}=+1$ puts $L$ in class AI$^{\dagger}$,
and the bands of a generic state are simple and meet only at exceptional
points.
For $n\equiv3\pmod4$, $\Gamma\Gamma^{*}=-1$ puts $L$ in class AII$^{\dagger}$,
whose Kramers theorem doubles every band.
For even $n$ the pencil is rectangular and $L$ obeys
$\Gamma L^{\mathsf T}\Gamma^{-1}=-L$, the particle-hole symmetry of classes D
and C, which pairs each band $w$ with $-w$ (End Matter~\ref{app:D}).

\textit{One surface.}---Carrying $\bm n$ once around an exceptional point
exchanges the two bands that meet there.
Carrying it twice around returns them with eigenvectors of opposite sign.
This sign is fixed by the square-root degeneracy and is the same for every
state.
The 24 exchanges generate all permutations of the four bands, so no band can
be followed on its own.
The four are sheets of one connected surface, as shown in the Supplemental
Material (SM)~\cite{SM}.
On a loop enclosing all exceptional points each band grows as $t^{2}$, and
the four bands wind around one another into the $(4,8)$ torus link for
every generic state [Fig.~\ref{fig:one}(c)].
Only the factorization of this braid into 24 half-twists, one at each
exceptional point, depends on the state.
The bands are the roots $w$ of $\det(w\,\id-L(u))=0$.
Together with their directions they form the curve
\begin{equation}
  C=\bigl\{(u,w)\ :\ \det\bigl(w\,\id-L(u)\bigr)=0\bigr\},
  \label{eq:curve}
\end{equation}
which covers the sphere $m$ times, with $m=4$ at five qubits and
$m=2^{(n-1)/2}$ for odd $n$.
Its branch points are the exceptional points, the zeros of the discriminant
$\prod_{i<j}(w_{i}-w_{j})^{2}$, which has degree $b=2m(m-1)$ in $u$.
For $n\equiv1\pmod4$ the Riemann--Hurwitz formula $2g-2=-2m+b$ for an
$m$-sheeted cover of the sphere then gives the genus $g=(m-1)^{2}$
whenever the discriminant is squarefree.
Squarefreeness is an open condition on the amplitudes and also makes $C$
smooth and irreducible (End Matter~\ref{app:A}).
At five qubits $C$ has genus nine.
The surface that the braid spans, four discs joined by 24 twisted strips,
has the same genus.
On $C$ the four bands are one single-valued function, and $C$ is the
spectral curve of a matrix polynomial of degree two on the
sphere~\cite{Beauville1990}.
Because $\det\gform=1$, the product of the bands is a perfect square,
\begin{equation}
  \det L(u)=\bigl(\det R(u)\bigr)^{2}.
  \label{eq:tangency}
\end{equation}
The curve $C$ therefore touches the level $w=0$ without crossing it, at the
directions where one coefficient of Eq.~\eqref{eq:bellschmidt} vanishes
(End Matter~\ref{app:B}).
These zeros of $\det R$, the data that the convex-roof constructions
use~\cite{Lohmayer2006,Osterloh2008}, are the intersection of $C$ with
$w=0$.
Local operations on qubits 1 to 4 leave $C$ unchanged.
One on the fifth qubit changes it only by a M\"obius map and a scale, so $C$
depends only on the class.
The characteristic polynomial has 24 coefficients, and Eq.~\eqref{eq:tangency}
removes four, M\"obius maps three and the scale one.
The remaining sixteen match the dimension $2^{5}-1-3\cdot5$ of the generic
five-qubit classes, and they are locally independent functions of the class
(End Matter~\ref{app:A}).
The curve thus fixes the class up to finitely many choices.

\textit{Eigenvector signs.}---On the sphere a polynomial is fixed by its
zeros up to a constant, which is why Majorana's points determine a symmetric
state.
On a surface of genus nine the curve alone does not fix the eigenvectors of
$L$, and the finitely many states that share a curve differ in them.
The eigenvectors are the frame vectors $e_{i}$ of
Eq.~\eqref{eq:bellschmidt}, the states of qubits 1 and 2 that carry each
coefficient.
The normalization $e^{\mathsf T}\gform e=1$ fixes each of them up to sign.
When $\bm n$ follows a closed path along which a band returns to itself, the
state that carries it comes back either unchanged or with its sign reversed.
Part of this sign is set by the bands alone.
It depends on how many times the product of the differences between that
band and the other three winds around zero along the path.
The rest is a $\mathbb{Z}_{2}$ Berry phase that depends only on the class
of the path on $C$.
The surface $C$ has eighteen independent closed paths, each with its own
Berry phase, and $L^{\mathsf T}\gform=\gform L$ guarantees that the sign is
all that can change (End Matter~\ref{app:C}).
The eigenvectors thus define a point of order two in the Jacobian of $C$,
the group of line bundles of degree zero on $C$.
The spectral data of Higgs bundles of split real forms are points of the
same kind~\cite{Schaposnik2013}.
The curve together with these signs fixes $L(u)$ up to a constant
conjugation~\cite{BNR1989}.
A sign defined along a closed path cannot be seen at any single direction.
The bands do not determine it either.
Interchanging the pair of qubits 1 and 2 with the pair 3 and 4 gives an
example.
The bands stay the same at every direction, yet the signs change, and no
constant matrix conjugates one $L(u)$ into the other.
Of the $2^{18}$ sign patterns, numerically about a quarter come from states.
Interchanging the two qubits within a pair leaves even the signs unchanged
and shows only in the sign of $\det R$.
Two states are compared in two steps.
Their band coefficients are compared first, and a linear test then decides
whether their matrices $L(u)$ are conjugate, with no search over the local
group
(End Matter~\ref{app:C}).

\textit{Symmetries split the bands.}---The bands of a generic state braid
into one surface.
The bands of the GHZ, $W$ and Dicke states do not, and their curve falls
apart into copies of the sphere (Fig.~\ref{fig:two}).
The splitting comes from symmetries.
A local symmetry that leaves the fifth qubit alone survives the projection,
so every conditional state carries it.
It acts on $L(u)$ at every direction without moving the direction, like an
internal symmetry of a Bloch Hamiltonian.
If $(X_{1}\otimes X_{2}\otimes X_{3}\otimes X_{4}\otimes1)|\Psi\rangle
=|\Psi\rangle$, then $L(u)=\delta\,ML(u)M^{-1}$ with $M=X_{1}\otimes X_{2}$,
where $\delta=\prod_{i}\det X_{i}$ is the factor by which the $X_{i}$ rescale
the spin-flip form $\gform$.
For $\delta=1$ and $M$ diagonalizable but not scalar, the eigenvalues of $M$
label the bands like a conserved quantum number, and bands with different
labels cross without mixing.
At these crossings $L$ stays diagonalizable, so they are not exceptional
points and the conditional state stays in $G_{abcd}$.
For $\delta=-1$, $M$ anticommutes with $L(u)$ and pairs each band $w$ with
$-w$, as a sublattice symmetry does.
A symmetry that also acts on the fifth qubit is a point-group operation on
the sphere and permutes the exceptional points.
Under $\sigma_{x}^{(1)}\sigma_{x}^{(5)}$, for instance, every band is a
constant multiple of
$1-t^{2}$, and all bands vanish at the two directions $t=\pm1$ that this
symmetry fixes.
The GHZ state gives the bands $0$ and $u_{0}u_{1}/2$, each twice, and the
bands of the Dicke state $|D^{2}_{5}\rangle$, with two excitations, do not
depend on $t$.
In these three cases the ratios of the bands do not depend on the direction,
so projecting the fifth qubit along almost any direction leaves the other
four in the same class.
For the $W$ state and for product states every band vanishes, because every
conditional state lies in the nullcone, where all invariants vanish.
The ring cluster state, absolutely maximally entangled~\cite{Helwig2012},
has the four separate bands $\pm8it^{2}$ and $\pm8i$, each a polynomial in
$t$.
Splitting without any local symmetry also occurs (SM).
A state symmetric under exchanging the two qubits of a pair has two equal
rows in $R$, so $\det R\equiv0$ and by Eq.~\eqref{eq:tangency} one band
vanishes over the whole sphere.
The construction therefore degenerates on the symmetric subspace, where the
Majorana constellation applies.
For states whose discriminant vanishes or has repeated factors, the classes
are told apart by the existing invariants~\cite{Burchardt2022}.

\textit{Observing the exceptional points.}---An exceptional point is seen
by carrying the measurement direction around it and watching two bands
exchange.
At each direction one measures the fifth qubit, keeps the outcome $+1$, and
reconstructs the conditional state by Pauli tomography.
Its bands follow up to a common factor, and the ratios
$w_{i}/\operatorname{tr}L$ depend neither on the normalization nor on the
global phase of the estimate.
The outcome $+1$ occurs with probability at least the smaller eigenvalue of
the reduced state of the fifth qubit, whatever the number of qubits.
We simulated shot noise for $\Psi_{0}$ on a loop of 32 directions around
one exceptional point.
With 80 shots per Pauli setting, about $4\times10^{5}$ shots in all, the
exchange is recovered in 96\% of runs.
A loop of the same size that encloses no exceptional point returns every
band in 99\% of runs.
Tomography of the five-qubit state itself gives the bands at every direction
from one data set, and $2.4\times10^{4}$ shots recover the exchange in 98\%
of runs.
The exchange depends only on whether the loop encloses the exceptional
point.
Any error that keeps that point inside the loop and the two bands apart
along it leaves the signal unchanged.
The square-root opening of the gap can be resolved as well, with more shots
(SM).

\textit{Discussion.}---The conditional state is linear in $u$, so every
polynomial invariant of it is holomorphic in $t$.
Its class is read from the spectrum of $L$, so the bands are roots of a
polynomial in $w$ and branch where two of them meet.
Exceptional points and a compact curve need nothing more.
Any group of local operations whose classes are read from such a matrix
gives them, and SLOCC is the case of entanglement.
Auxiliary, elliptic and secant varieties attach geometry to the classes
themselves~\cite{Levay2011,Holweck2012,Sanz2017,Gharahi2020}, whereas this
curve lies over the directions of a measurement.
Expectation values contain the conjugate amplitudes as well, so they are not
holomorphic in $t$ and give none of this structure.
The one-qubit spectra behind entanglement
polytopes~\cite{Walter2013,Sawicki2014} are built from such expectation
values.
A symmetric state avoids the nullcone exactly when no more than half of
Majorana's points coincide.
Here a state avoids it whenever its discriminant does not vanish
identically and no direction carries more than half of its zeros (SM).
From seven qubits on, the curve is unchanged by operations that respect the
time reversal $\sigma_{y}^{\otimes k}K$ of each block of $k$ qubits, with $K$
complex conjugation.
These operations can entangle the qubits within a block, so one curve cannot
separate the classes (End Matter~\ref{app:D}).
Whether the curves of all projected qubits and splits together separate them
is open.
The bands cost one eigenvalue problem of size $m$ per direction.
With Schmidt rank $r$ across the split at most $r$ bands are nonzero, so for
matrix-product states of bounded bond dimension the curve stays bounded as
$n$ grows and follows from the tensors (SM).
Device noise splits the Jordan block as shot noise does and limits the loop
test in the same way (SM).
For general states, Majorana's points thus become a band structure whose
exceptional points mark where the entanglement of the remaining qubits
changes family.

\textit{Data availability.}---The scripts that produce every number and
figure in this Letter, together with the exact-arithmetic certificates and
the integer amplitudes of seed 20260909, are openly available on
Zenodo~\cite{KamZenodo2026}.

\bibliographystyle{apsrev4-2}
\bibliography{main_spectral_prl}

\onecolumngrid
\setcounter{secnumdepth}{1}
\clearpage
\appendix
\section*{End Matter}
\twocolumngrid

\section{Covariance, genus and parameter count}
\label{app:A}

For $X$ with $X^{\mathsf T}\gform X=\gform$ one has
$X^{-1}=\gform X^{\mathsf T}\gform$.
Hence $R\mapsto XR$ sends $L=R\gform R^{\mathsf T}\gform$ to $XLX^{-1}$, and
$R\mapsto RY^{\mathsf T}$ with $Y^{\mathsf T}\gform Y=\gform$ leaves $L$ fixed.
Local operations of unit determinant on either pair satisfy this condition,
because $\sigma_{y}$ is the invariant form of $\SL(2)$.
An operation $Z$ on the fifth qubit replaces $u$ by $Z^{\mathsf T}u$, so it
moves the directions by a M\"obius map and, $L$ being quadratic in $u$,
multiplies all bands by a common factor.

Choosing $T$ with $T^{\mathsf T}\gform T=\id$ gives
$L=\tilde R\tilde R^{\mathsf T}$ for $\tilde R=T^{-1}R\,T^{-\mathsf T}$.
The complex-orthogonal singular value decomposition
$\tilde R=O_{1}\mathrm{diag}(a,b,c,d)O_{2}^{\mathsf T}$ is fixed up to
permutations and an even number of sign changes, a group of order $192$ that
moves $\{a,b,c,d\}$ but never $\{a^{2},b^{2},c^{2},d^{2}\}$.
Flipping the same column of $O_{1}$ and $O_{2}$ leaves $\tilde R$ unchanged,
so only the product of their determinants is fixed.
Its sign is the $\mathbb{Z}_{2}$ by which $\mathrm{O}(4,\CC)$ exceeds
$\SO(4,\CC)$, and exchanging the two qubits of a pair flips it
(End Matter~\ref{app:C}).

Write the characteristic polynomial as
$\chi(w)=\det(w\,\id-L)=\sum_{j=0}^{m}(-1)^{j}\chi_{j}\,w^{m-j}$, so that
$\chi_{j}$ is the $j$th elementary symmetric function of the bands.
Since $L$ is quadratic in $u$, $\chi_{j}$ is homogeneous of degree $2j$.
Each factor $(w_{i}-w_{j})^{2}$ of the discriminant has degree four in $u$,
so the discriminant has the degree $b=2m(m-1)$ used in the text.
A factorization of $\chi$ into factors with $m_{1}$ and $m_{2}$ sheets would
make the discriminant divisible by the square of their resultant.
A squarefree discriminant of full degree therefore forces $C$ to be
irreducible, and it makes every branch point simple and $C$ smooth.
For $\Psi_{0}$ the discriminant is squarefree of degree 24 in exact
arithmetic.

At five qubits Eq.~\eqref{eq:tangency} leaves 20 of the 24 coefficients of
$\chi$ independent.
The differential of the map from the 32 amplitudes to these 20
coefficients, computed exactly at $\Psi_{0}$, has rank 20, and its kernel is
the $\mathfrak{sl}(2)^{\oplus4}$ of qubits 1 to 4.
The image therefore has all 20 directions, three of which are M\"obius maps
and one the scale.
The remaining sixteen coefficients are locally independent functions of the
class, so the map from classes to curves has finite generic fibers, which is
the finiteness used in the text.

\section{The band product and the four-qubit invariants}
\label{app:B}

Since $\det\gform=1$,
$\det L=\det R\,\det\gform\,\det R^{\mathsf T}\det\gform=(\det R)^{2}$, which
is Eq.~\eqref{eq:tangency}.
For a generic state $\det R$ has simple zeros, and $w=0$ is a simple
eigenvalue there.
The intersection of $C$ with the level $w=0$ is then twice the zero set
$\mathcal{Z}$ of $\det R$, so $C$ is tangent to that level at every point of
$\mathcal{Z}$, the
directions at which one coefficient of Eq.~\eqref{eq:bellschmidt} vanishes.

Luque and Thibon showed that the polynomial invariants of four qubits are
generated by a quadratic $H$, two quartics $\mathsf{L}$ and $\mathsf{M}$, and
a sextic $\mathsf{D}$~\cite{LuqueThibon2003}, written here in sans-serif to
distinguish them from the matrix $L$.
Computing them from their definitions in Ref.~\cite{LuqueThibon2003} gives
\begin{equation}
  \begin{aligned}
  \chi_{1}&=2H, & \chi_{2}&=H^{2}+2\mathsf{L}+4\mathsf{M},\\
  \chi_{3}&=2H\mathsf{L}+4\mathsf{D}, & \chi_{4}&=\mathsf{L}^{2},
  \end{aligned}
  \label{eq:LTem}
\end{equation}
as polynomial identities in the sixteen amplitudes, checked in exact
arithmetic.
With the third quartic $\mathsf{N}=-\mathsf{L}-\mathsf{M}$ of
Ref.~\cite{LuqueThibon2003} in place of $\mathsf{M}$, the second identity
fails.
The four bands at one direction therefore carry the complete set of
polynomial invariants of the conditional state, except for the sign of
$\mathsf{L}$, which is the $\det R$ of Eq.~\eqref{eq:tangency} at that
direction.

\section{Eigenvector signs and the equivalence of states}
\label{app:C}

\textit{Signs.}---Because $L^{\mathsf T}\gform=\gform L$, the left
eigenvector belonging to $e$ is $\gform e$, and at a band $w$
\begin{equation}
  \mathrm{adj}(w-L)=\chi'(w)\,e\,e^{\mathsf T}\gform ,
  \label{eq:adj}
\end{equation}
where $\chi'(w)$, the derivative of the characteristic polynomial, is the
product of the differences between $w$ and the other bands.
Since the exceptional points are of second order, $w-L$ has rank three
everywhere on $C$ and its adjugate never vanishes there.
Writing $\mathrm{adj}(w-L)=v\,v^{\mathsf T}\gform$ defines a vector $v$
spanning the eigenline bundle $\mathcal{L}$ of $L$ over $C$, up to sign, and
$e=v/\sqrt{\chi'(w)}$.
On $C$ the function $\chi'(w)$ vanishes exactly at the 24 points above the
exceptional points, each simply.
Along a closed path the sign of $e$ is therefore that of $v$ times $-1$ to
the winding number of $\chi'(w)$ (SM).
The signs of $v$ are the holonomies of $\mathcal{L}$.
The adjugate has degree six in $u$, so it is a nowhere-zero section of
$\mathcal{L}^{\otimes2}\otimes\pi^{*}\mathcal{O}(6)$, where $\pi$ is the
projection of $C$ to the sphere.
Hence $\mathcal{L}^{\otimes2}\cong\pi^{*}\mathcal{O}(-6)$ and
\begin{equation}
  \mathcal{L}\otimes\pi^{*}\mathcal{O}(3)\in\Jac(C)[2]\cong(\mathbb{Z}/2)^{2g},
  \label{eq:torsion}
\end{equation}
where $\Jac(C)[2]$ is the group of line bundles on $C$ whose square is
trivial, the two-torsion of the Jacobian.
These are flat bundles with holonomy $\pm1$ around each of the $2g=18$
independent cycles.
The resultant of $(\gform\,\mathrm{adj}(w-L))_{11}$ with $\chi$, a perfect
square of degree 24 in $t$, confirms $\deg\mathcal{L}=-12$.

\textit{Exchanges.}---Exchanging the pairs sends $R$ to $R^{\mathsf T}$.
Then $v'=R^{\mathsf T}\gform v$ is an eigenvector of
$L'=R^{\mathsf T}\gform R\gform$ with the same eigenvalue, and it vanishes
exactly on the zero set $\mathcal{Z}$ of $\det R$
(End Matter~\ref{app:B}).
Hence $\mathcal{L}'\cong\mathcal{L}\otimes\eta_{0}$ with
$\eta_{0}=\mathcal{O}_{C}(\mathcal{Z})\otimes\pi^{*}\mathcal{O}(-1)$, where
$\mathcal{O}_{C}(\mathcal{Z})$ is the line bundle with a section vanishing
exactly on $\mathcal{Z}$, and $\eta_{0}$ is two-torsion by
Eq.~\eqref{eq:tangency}.
Exchanging the two qubits of a pair is $R\mapsto PR$ with $P$ the swap, which
obeys $P^{\mathsf T}\gform P=\gform$ and $\det P=-1$.
It conjugates $L$ by a constant, so curve and bundle are unchanged, and it
flips the sign of $\det R$.
States inequivalent to the original but with the same curve and the same
bundle therefore exist, related by these exchanges.

\textit{Counting.}---For $\Psi_{0}$, Newton's method from $2\times10^{5}$
random starts on the twenty equations $\chi_{R'}=\chi_{R}$, $\det R'=\det R$
gives about $1.2\times10^{5}$ orbits of $\SO(4,\CC)\times\SO(4,\CC)$.
They come in pairs related by exchanging both pairs of qubits internally,
hence about
$6\times10^{4}$ bundles, near a quarter of the $2^{18}$ allowed by
Eq.~\eqref{eq:torsion}.
For a $3\times3$ pencil with $L=RR^{\mathsf T}$, a curve of genus four, the
count saturates at exactly $72=2^{g-1}(2^{g-1}+1)$.
Seventy-two is the number of theta characteristics $\kappa$, the square roots of
the canonical bundle of the curve, with $\kappa$ and $\kappa\otimes\eta_{0}$
both even.
The same formula gives $65{,}792$ at genus nine, consistent with the
five-qubit count.

\textit{Deciding equivalence.}---After a M\"obius map brings the directions of two
states with matching band coefficients to a common normal form, the states
are compared by solving $XL_{j}=\hat L_{j}X$ for the coefficients of
$L(u)=L_{0}+tL_{1}+t^{2}L_{2}$ and of $\hat L(u)$, built from the pencil
$\hat R$ of the second state, a linear system of
$3m^{2}$ equations in $m^{2}$ unknowns.
Its nullity is one when $L(u)$ and $\hat L(u)$ are conjugate and zero
otherwise, because a matrix commuting with $L(u)$ at every direction is scalar
when $C$ is irreducible.
The conjugator $X$ is complex orthogonal up to scale, and
$\hat R=XRY^{\mathsf T}$ then fixes $Y$, which is constant for a generic
state (SM).
A local operation on a pair has rearrangement rank, the operator Schmidt
rank across its two qubits, equal to one, and the swap of the two qubits has
rank four, so the rearrangement ranks of $X$ and $Y$ separate
local operations from exchanges within a pair.

\section{Other numbers of qubits}
\label{app:D}

Odd $n$ makes $R$ square, since the $n-1$ other qubits split into two blocks
of $k=(n-1)/2$ qubits and $m=2^{k}$.
Even $n$ makes $R$ rectangular, so $\det R$ and the tangency of
Eq.~\eqref{eq:tangency} are absent, and the spectrum of $L$ is closed under
$w\mapsto-w$, as the particle-hole symmetry of the main text requires
(checked at $n=4,6,8$).
For odd $n$ the form $\gform=\sigma_{y}^{\otimes k}$ is symmetric for even
$k$ and antisymmetric for odd $k$, and the local group of a block lies in
$\mathrm{O}(m,\CC)$ or in $\mathrm{Sp}(m,\CC)$ accordingly.
For $n\equiv1\pmod4$ the bands of a generic state are distinct.
For $n\equiv3\pmod4$ a simple band would need an eigenvector $v$ with
$v^{\mathsf T}\gform v\neq0$, but this is the Wong--Christensen concurrence
amplitude of an odd number of qubits, which vanishes
identically~\cite{WongChristensen2001}.
Every eigenvalue is therefore doubly degenerate, the Kramers degeneracy of
class AII$^{\dagger}$, and $\chi=p^{2}$ for a polynomial $p$ of degree
$m'=m/2$ in $w$, which we verified at $n=7$ modulo a prime.
The curve is then $p=0$, with $m'$ sheets.
The constant term of $p$ is $\pm\det R$, which has simple zeros on a generic
state, so this curve crosses the level $w=0$ instead of touching it, and the
tangency of Eq.~\eqref{eq:tangency} and the class $\eta_{0}$ belong to the
first family alone.
The curve has at most $2^{n-1}+2^{(n-1)/2}-4$ or
$2^{n-3}+2^{(n-1)/2}-4$ parameters in the two families, against
$2^{n}-1-3n$ (SM), and five is the only odd number above three at which
the two agree.
For $n\equiv1\pmod4$ the difference has a geometric meaning.
A pencil is fixed by its curve, its eigenvector signs and two constant
frames in $\mathrm{O}(m,\CC)$, one for each block~\cite{BNR1989}, and
\begin{equation}
  2^{n}-3n-1=\bigl(m^{2}+m-4\bigr)+2\Bigl[\tfrac{m(m-1)}{2}-3k\Bigr]
  \label{eq:split}
\end{equation}
assigns the remainder to the frames modulo the local group of each block.
Ranks of the corresponding maps confirm this split at $n=5$ and $9$
(SM).
The curve records the entanglement across the split, and the frames record
the entanglement within each block.
At five qubits each block is a pair, whose local group
$\SO(4,\CC)\cong(\SL(2,\CC)\times\SL(2,\CC))/\mathbb{Z}_{2}$ moves any frame to
any other.
The frames then carry nothing, and the curve is complete.
More generally the curve is unchanged by any operation $X$ on a block with
$X^{\mathsf T}\gform X=\gform$, since $L\mapsto XLX^{-1}$.
For a unitary this means commuting with the time reversal
$\sigma_{y}^{\otimes k}K$ of the block, with $K$ complex conjugation.
Such unitaries are local for a pair, up to exchanging its two qubits, and
entangling for larger blocks (SM).
For $n\equiv1\pmod4$ the curve thus classifies states, up to at most
$2^{2g}$ choices with $g=(m-1)^{2}$, under operations that respect the
time reversal of each block (SM).
For $n\equiv3\pmod4$ the doubled eigenspaces carry further continuous data
(SM).
For Gaussian-integer test states the discriminants are squarefree of full
degree, in exact arithmetic for $n=5$ and modulo a prime
$\ell\equiv1\pmod4$ for $n=7,9,11$, which implies the same over $\CC$
(SM).

\end{document}